\documentclass{ws-ijmpb}

\begin{document}

\markboth{Z. Y. Ou} {Multi-Photon Interference and Temporal
Distinguishability of Photons}

%
\catchline{}{}{}{}{}
%

\title{Multi-Photon Interference and Temporal Distinguishability of Photons}

\author{Z. Y. Ou}

\address{Department of Physics, Indiana University-Purdue University Indianapolis\\
402 N. Blackford Street, Indianapolis, IN 46202, USA\\
zou@iupui.edu}

\maketitle

\begin{history}
\received{\today}
\end{history}

\begin{abstract}
A number of recent interference experiments involving multiple
photons are reviewed. These experiments include generalized photon
bunching effects, generalized Hong-Ou-Mandel interference effects
and multi-photon interferometry for demonstrations of multi-photon
de Broglie wavelength. The multi-photon states used in these
experiments are from two pairs of photons in parametric
down-conversion. We find that the size of the interference effect
in these experiments, characterized by the visibility of
interference pattern, is governed by the degree of
distinguishability among different pairs of photons. Based on this
discovery, we generalize the concept of multi-photon temporal
distinguishability and relate it to a number of multi-photon
interference effects. Finally, we make an attempt to interpret the
coherence theory by the multi-photon interference via the concept
of temporal distinguishability of photons.
\end{abstract}


\section{Introduction and Historic Background}

Interference of light, as a wave phenomenon, played a pivotal role
in our understanding of light, from the early establishment of the
classical wave theory\cite{hui} to the formation of the modern
coherence theory\cite{bw}. Coming to the quantum age, it first
provided a platform for conceptual understanding of quantum
physics\cite{dirac} and then leads to fundamental test of quantum
mechanics\cite{ou88-1}. More recently, it is associated with
quantum information processing\cite{zei}.

The most commonly-encountered interference phenomena are in the
form of some beautiful interference fringes that have been
well-studied by the classical coherence theory in Born and Wolf's
classic book {\it Principle of Optics}\cite{bw}. In terms of the
language of photon, these phenomena can be categorized as the
single-photon interference effect and described by the famous
Dirac's statement\cite{dirac}: \vskip 0.1in
 \centerline{\it Each photon ... only interferes with itself.}
 \centerline{\it Interference between different photons never occurs.}
 \vskip 0.1in

However, with recent boom in quantum information science,
interference with two and more photons came into focus. It plays
an essential role in some quantum information protocols.

\subsection{Early years of two-photon interference}
The history of two-photon interference started at Pfleegor-Mandel
experiment\cite{pfl}. After a dramatic demonstration of the
interference effect between two independent lasers by Magyar and
Mandel\cite{magy} in 1963, a dark cloud was casted on the second
part of the Dirac statement\cite{dirac}: is it true that no
interference between different photons occurs? A true test would
be to repeat the Magyar-Mandel experiment at single-photon level.
This requires long time exposure. But it is well known that the
phase of a laser diffuses in a long time period so that the
interference fringes would be washed out. Under this circumstance,
Pfleegor and Mandel\cite{pfl} designed an ingenious experiment to
reveal the interference between independent lasers at
single-photon level.

To overcome the problem of phase drift in long term, Pfleegor and
Mandel employed the technique of intensity correlation between two
detectors at separate locations. They discovered that there is a
positive correlation when the detectors are one full fringe
spacing apart but a negative correlation for half a fringe spacing
separation, thus revealing the interference effect between
independent lasers at single-photon level.  This turns out to be
the first two-photon interference phenomenon.

The Pfleegor-Mandel experiment can be easily explained in terms of
classical wave theory. As is well-known, laser is the closest to a
monochromatic wave with a phase that diffuses in a time period
longer than coherence time. So if the observation time is short,
as in the Magyar-Mandel experiment, both lasers have a steady
phase even though they are operated independently, which gives
rise to a steady interference fringe pattern observed by Magyar
and Mandel:
\begin{equation}
I(x) = I_0[1+\cos (2\pi x/L+\Delta\varphi)], \label{1.1}
\end{equation}
where $L$ is the fringe spacing and $\Delta\varphi$ is the phase
difference between the phases of the two lasers. On the other
hand, when the observation time is long, as required in
Pfleegor-Mandel experiment for acquiring data at single-photon
level, the phases of the two independent lasers will diffuse
independently leading to a random phase difference
$\Delta\varphi$. Thus intensity distribution will not show
interference fringe. But for intensity correlation between two
detectors at two locations, we have from Eq.(\ref{1.1})
\begin{equation}
\langle I(x_1)I(x_2)\rangle_{\Delta\varphi} = I_0^2[1+0.5\cos 2\pi
(x_1-x_2)/L]. \label{1.2}
\end{equation}
Immediately, we have
\begin{equation}
\lambda_{12}\equiv \langle \Delta I(x_1)\Delta
I(x_2)\rangle_{\Delta\varphi}/I_0^2 = 0.5\cos 2\pi (x_1-x_2)/L,
\label{1.3}
\end{equation}
which equals $+0.5$ for $x_1-x_2 =L$ and $-0.5$ for $x_1-x_2
=L/2$, consistent with the experimental results. Detailed study
later confirmed Eq.(\ref{1.3}).

But how can we understand the Pfleegor-Mandel experiment in terms
of photons? Before we come to that, let us first examine the
difference between intensity measurement and intensity correlation
measurement. As seen in Fig.1, intensity measurement uses one
detector and responses whenever there is one photon present
whereas intensity correlation relies on two detectors and gives
rise to a signal only when both detectors fire within the certain
time window. So the intensity correlation requires the presence of
two photons. In this way, we find that intensity measurement
corresponds to one-photon detection and intensity correlation to
two-photon detection.
\begin{figure}[bt]
\centerline{\psfig{file=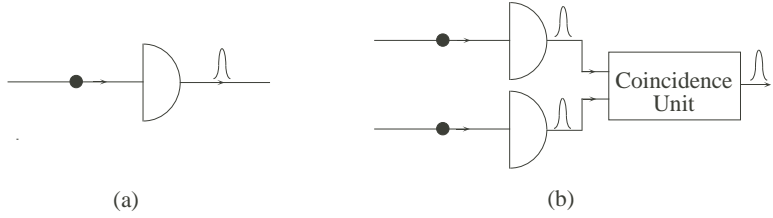,width=4.in}} \vspace*{8pt}
\caption{The photo-detection processes of intensity measurement
(a) and intensity correlation measurement (b).}
\end{figure}

Therefore Pfleegor-Mandel experiment is an interference effect
involving two photons, for which Dirac's statement on photon
interference no longer applies. We need to generalize the Dirac
statement to two-photon case as \vskip 0.1in
 \centerline{\it A pair of photons only interferes with the pair itself in two-photon interference.}

 \vskip 0.1in
With this generalized statement, let's see how we can understand
Pfleegor-Mandel experiment quantitatively in terms of photons. As
discussed before, intensity correlation measurement involves two
photons. The two detected photons have four possible ways to come
from two lasers, as shown in Fig.2. Because of random phases of
the two lasers, cases A and B give rise to no interference (If
there were a steady phase difference between the two lasers, cases
A and B would produce a phase dependent interference pattern).
Using the generalized Dirac statement, we find that cases C and D
will produce an interference fringe of $1+\cos2\pi(x_1-x_2)/L$.
Because of the randomness for the photons from a lasers, the four
possibilities in Fig.2 have the same probability, assigned as
$P_{20}$. Then the coincidence rate in the intensity correlation
measurement is
\begin{eqnarray}
R_2(x_1,x_2) &\propto
&P_{20}+P_{20}+2P_{20}[1+\cos2\pi(x_1-x_2)/L] \cr &=&
4P_{20}[1+0.5\cos 2\pi (x_1-x_2)/L], \label{1.4}
\end{eqnarray}
which is in the exactly same form as Eq.(\ref{1.2}). This
indicates that the generalized Dirac statement indeed gives the
correct prediction for Pfleegor-Mandel experiment. As a matter of
fact, the generalized Dirac statement leads to interference
between two wave amplitudes of two photons, i.e., two-photon
waves. This picture of two-photon wave is indeed confirmed by the
more rigorous quantum theory of light. The Dirac statement for two
photons can be further generalized to arbitrary $N$ photons with
the concept of $N$-photon waves.

\begin{figure}[bt]
\centerline{\psfig{file=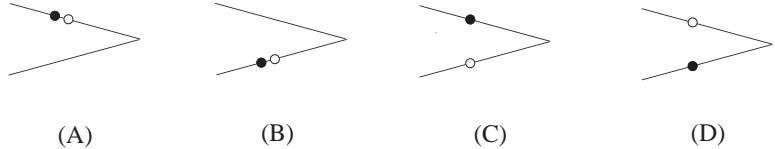,width=4.in}} \vspace*{8pt}
\caption{Four possible ways for the two detected photons from two
lasers.}
\end{figure}

Notice that the visibility in Pfleegor-Mandel experiment is only
50\%, as shown in Eqs.(\ref{1.2}, \ref{1.4}). This is because of
the existence of cases A and B. In fact, for all classical sources
of light, the probabilities for cases A and B are always larger
than cases C and D, resulting in maximum visibility of 50\% for
classical fields. This is first pointed out by Mandel\cite{man83}
in 1983 and generally proved by Ou\cite{ou88-2} in 1988.

To obtain a visibility larger than 50\% in two-photon
interference, nonclassical fields with anti-bunching effect must
be employed. Specifically with single-photon states in the two
fields, the probabilities of cases A and B are zero, which leads
to 100\% visibility in two-photon interference. Indeed, more than
50\% visibility in two-photon interference was observed by Ghosh
and Mandel\cite{gho} and by Ou and Mandel\cite{ou89} with
two-photon state from parametric down-conversion.

Next, we consider another two-photon interference effect with
two-photon state.

\subsection{Hong-Ou-Mandel effect}

Hong-Ou-Mandel effect\cite{hom} is the most famous two-photon
interference effect that has been widely applied in many fields.
It is exploited in quantum information sciences and serves mainly
as the fundamental process in the scheme of linear optical quantum
computing\cite{klm}.

A Hong-Ou-Mandel interferometer is made of a 50:50 beam splitter
and two photons which enter the beam splitter from the opposite
sides (Fig.3). For the two photons entering the beam splitter,
there are four possible ways for them to come out. In Figs.3a and
3b, both photons go to either side of the beam splitter together.
In Figs.3c and 3d, the two photons are either both transmitted of
reflected, resulting them exit at separate ports. For the latter
two cases, there is no way to tell them apart at the outputs.
Therefore, quantum interference will occur. A detailed study of
the phases shows that energy conservation in a lossless beam
splitter leads to a 180$^{\circ}$ phase difference between the two
cases\cite{ohm89}. Since the amplitudes are the same in the two
cases for a 50:50 beam splitter, there is a complete destructive
interference between the two cases. So the probability is zero for
the two photons exit at two separate sides. This is the
Hong-Ou-Mandel effect.
\begin{figure}[bt]
\centerline{\psfig{file=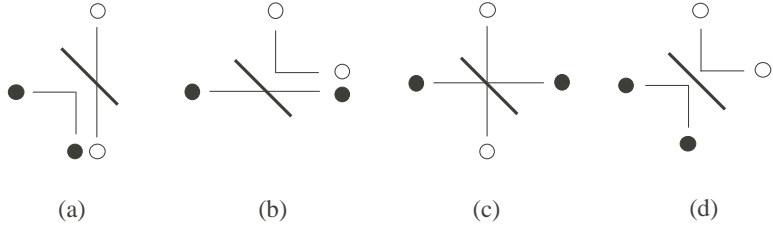,width=4.in}} \vspace*{8pt}
\caption{Four possible ways out for two photons entering a 50:50
beam splitter from two sides.}
\end{figure}

Mathematically, we can calculate the probability of detecting two
photons at two output ports of the beam splitter by finding the
quantum state at the output. We may follow the procedure in
Refs.\refcite{ohm87} and \refcite{cam} or more simply as follows.
It is well known that the input-output relation in Heisenberg
picture is given by
\begin{eqnarray}
\hat b_1&=&\sqrt{T}\hat a_1-\sqrt{R}\hat a_2,\cr \hat
b_2&=&\sqrt{T}\hat a_2+\sqrt{R}\hat a_1.\label{5}
\end{eqnarray}
where $T, R$ are the transmissivity and reflectivity of the beam
splitter, respectively. Or we may write in terms of the evolution
operator $\hat U$ as
\begin{eqnarray}
\hat b_1 =\hat U^{\dag}\hat a_1\hat U,~~ \hat b_2 =\hat
U^{\dag}\hat a_2\hat U.\label{6}
\end{eqnarray}
In Schr\"odinger picture, the output state is then
\begin{eqnarray}
|\Phi_{out}\rangle =\hat U |\Phi_{in}\rangle.\label{7}
\end{eqnarray}
With an input state of $|\Phi_{in}\rangle = |1_1, 1_2\rangle =
\hat a_1^{\dag}\hat a_2^{\dag}|0\rangle$, we have
\begin{eqnarray}
|\Phi_{out}\rangle &=&\hat U \hat a_1^{\dag}\hat
a_2^{\dag}|0\rangle= \hat U \hat a_1^{\dag}\hat a_2^{\dag}\hat
U^{\dag}\hat U|0\rangle\cr &=& \hat U \hat b_1^{\dag}\hat
b_2^{\dag}\hat U^{\dag}\hat U|0\rangle = (\hat U \hat
b_1^{\dag}\hat U^{\dag})(\hat U\hat b_2^{\dag}\hat U^{\dag})\hat
U|0\rangle,\label{8}
\end{eqnarray}
where we replaced $\hat a$'s with $\hat b$'s due to Schr\"odinger
picture. By Eq.(\ref{6}), we have
\begin{eqnarray}
\hat a_1 =\hat U\hat b_1\hat U^{\dag},~~ \hat a_2 =\hat U\hat
b_2\hat U^{\dag}.\label{9}
\end{eqnarray}
But from the reverse of Eq.(\ref{5}), we then obtain
\begin{eqnarray}
\hat U\hat b_1\hat U^{\dag}&=&\sqrt{T}~\hat b_1+\sqrt{R}~\hat
b_2,\cr \hat U\hat b_2\hat U^{\dag}&=&\sqrt{T}~\hat
b_2-\sqrt{R}~\hat b_1.\label{10}
\end{eqnarray}
Furthermore, it is obvious that we have $\hat U |0\rangle
=|0\rangle$, i.e., vacuum input leads to vacuum output.
Substituting the above into Eq.(\ref{8}), we finally obtain the
output state as
\begin{eqnarray}
|\Phi_{out}\rangle &=&(\sqrt{T}~\hat b_1^{\dag}+\sqrt{R}~\hat
b_2^{\dag})(\sqrt{T}~\hat b_2^{\dag}-\sqrt{R}~\hat
b_1^{\dag})|0\rangle \cr &=& (T-R)|1_1,1_2\rangle -
\sqrt{2TR}(|2_1,0_2\rangle-|0_1,2_2\rangle).\label{11}
\end{eqnarray}
Therefore, we have $P_{2}(1,1)=(T-R)^2$, which is zero for a 50:50
beam splitter, as we discussed according to the pictures in Fig.3.

The above analysis is for the single mode case. In practice,
however, the fields are all in multiple modes. Under this
circumstance, photons are described by wave packets. The
interference effect described earlier only occurs when the two
wave packets for the two photons coincide exactly at the beam
splitter. On the other hand, when the two wave packets are
well-separated arriving at the beam splitter, no interference
occurs and we have a non-zero probability $P_2(1,1)$ for detecting
two photons at the opposite sides of the beam splitter. So as we
scan the relative delay between the two photon wave packets, the
probability $P_2(1,1)$ will exhibit a drop and reach all the way
to zero when the delay is zero. This is the Hong-Ou-Mandel dip.
Fig.4 shows the set-up and the result.

\begin{figure}[bt]
\centerline{\psfig{file=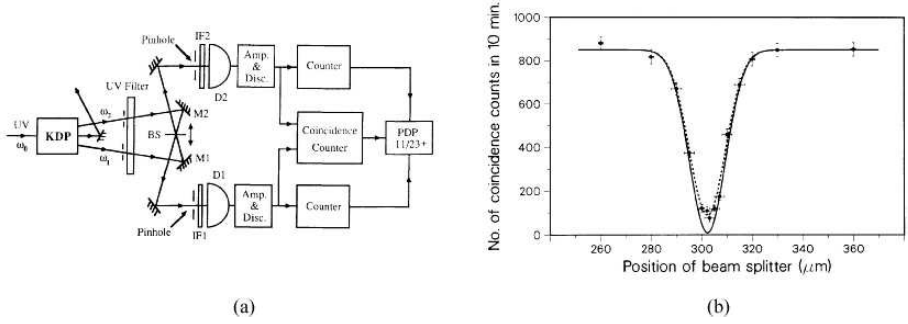,width=4.8in}} \vspace*{8pt}
\caption{Hong-Ou-Mandel interference effect: (a) outline of setup
(b) the result of the experiment. Reproduced with permission from
Ref. 12. }
\end{figure}

When the two wave packets are well separated, the two photons
behave independently as classical particles. So the overall
probability of detecting two photons at the opposite sides is
simply the sum of the probabilities of cases (c) and (d) in Fig.3.
Then we easily obtain the classical probability $P_2(1,1)$ as
\begin{eqnarray}
P_2^{cl}(1,1) = T^2+R^2.\label{12}
\end{eqnarray}
This corresponds to the flat lines on the two sides of the dip in
Fig.4b.

\subsection{Photon bunching and two-photon interference}

Photon bunching effect, first discovered by Hanbury-Brown and
Twiss\cite{hbt} in 1956, was the starting point for the field of
quantum optics. The attempt to explain it leads to the development
of the techniques in quantum optics.

Although it is a purely classical wave effect from Gaussian
statistics of a thermal source and does not need quantum theory of
light to understand it, explanation in terms of photons does need
some imagination. It has long been argued that photon bunching
effect is caused by Bosonian nature of photon\cite{pur}. But this
argument was soon overwhelmed by the success of the explanation by
classical coherence theory\cite{msw}. Later on, in his monumental
work on quantum coherence, Glauber\cite{gla} briefly discussed the
connection between photon bunching effect and the interference of
two-photon amplitudes, which is equivalent to our language of
two-photon interference.

But to fully understand the photon bunching effect in terms of
two-photon interference, we go back to the Hong-Ou-Mandel
interferometer. Instead of the probability of $P_2(1,1)$, we are
interested in $P_2(2,0)$, the probability of both photons exit in
the same port. From Eq.(\ref{11}), we have
\begin{eqnarray}
P_2^{qu}(2,0) = 2TR.\label{13}
\end{eqnarray}
On the other hand, we may find the classical probability of
detecting both photons at the same side of the beam splitter,
i.e., case (a) or (b) in Fig.3. The result is simply the product
of single-photon events due to independence:
\begin{eqnarray}
P_2^{cl}(2,0) = P_2^{cl}(0,2) = P_1(1,0)P_1'(1,0)=TR,\label{14}
\end{eqnarray}
where $P_1(')(1,0)$ is the probability for one and the other input
photons. From Eqs.(\ref{13}, \ref{14}), we find that the quantum
probability is twice the classical probability. This is exactly
the same as the photon bunching effect. Its demonstration was
first performed by Rarity and Tapster\cite{rar}, as shown in
Fig.5.

\begin{figure}[bt]
\centerline{\psfig{file=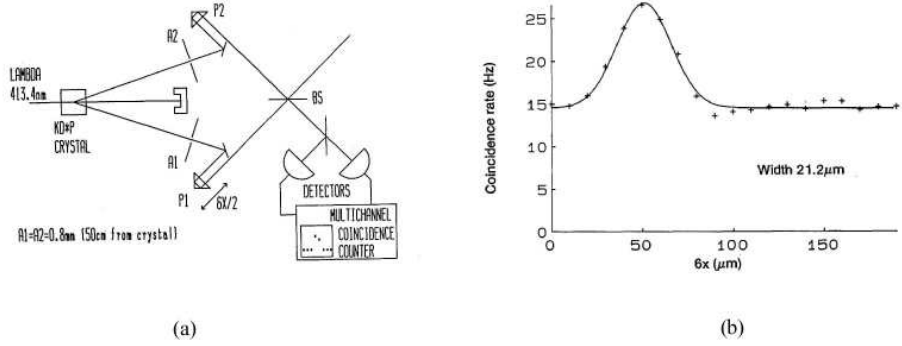,width=4.8in}} \vspace*{8pt}
\caption{Photon bunching in Hong-Ou-Mandel interferometer: (a)
outline of setup (b) the result of the experiment. Reproduced with
permission from Ref. 21.}
\end{figure}

To see that this is the result of two-photon interference, we
consider Fig.6, which is the setup to measure $P_2(2,0)$. For
two-photon detection, there are two possible ways to arrange the
two photons (Figs.6a,6b). If the incoming two photons are well
separated, they behave like classical particles and we add the
probabilities of the two possibilities: $P_2^{cl}=|A|^2+|A|^2$.
But if they are overlap at the beam splitter, we cannot
distinguish the two possibilities and we add the amplitudes before
taking the absolute value for overall probability: $P_2^{qu}
=|A+A|^2 =4|A|^2 =2P_2^{cl}(2,0)$. Note that the phases for the
two cases are the same because the overall paths for the two
photons in the two possibilities are the same due
indistinguishability of the two photons. Therefore, it is
constructive that is responsible for the photon bunching effect in
Hong-Ou-Mandel interferometer. Scarcelli {\it et al.}\cite{shi}
recently showed that it is also the fundamental principle behind
the original photon bunching effect from a thermal source. This
was consistent with the original view by Glauber\cite{gla} who
first explained photon bunching effect with an equivalent view of
two-photon amplitudes.

\begin{figure}[bt]
\centerline{\psfig{file=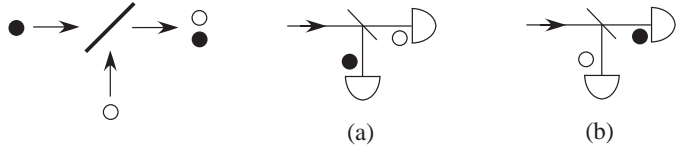,width=3.5in}} \vspace*{8pt}
\caption{Two possibilities for two-photon constructive
interference in explaining photon bunching in Hong-Ou-Mandel
interferometer.}
\end{figure}

\section{Generalized Photon Bunching Effect and Constructive Multi-Photon Interference}

The two-photon bunching effect discussed in the previous section
can be generalized to the case of more photons.

\begin{figure}[bt]
\centerline{\psfig{file=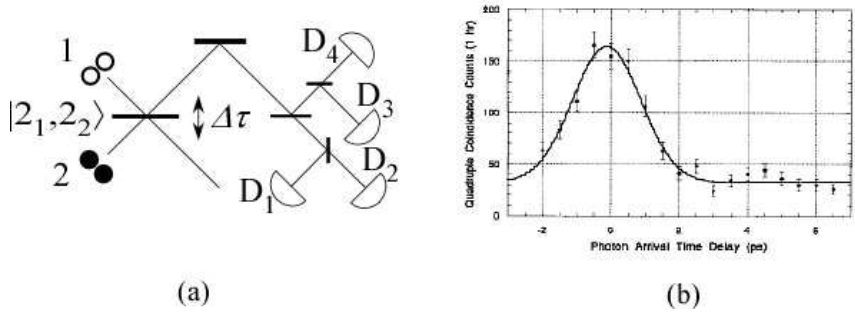,width=4.5in}} \vspace*{8pt}
\caption{(a) Schematics for demonstrating photon pair bunching
effect. (b)The result of photon pair bunching; reproduced with
permission from Ref.[23].}
\end{figure}

\subsection{Pair bunching effect}

The first generalization is the pair bunching effect. In this
case, two pairs of photons enter in two separate ports of a 50:50
beam splitter, as shown in Fig.7a. This scheme is similar to the
Hong-Ou-Mandel interferometer of two photons but there are four
photons here and unlike its predecessor, the middle term
$|2,2\rangle$ does not vanish, as seen in the output state derived
in the same way as Eq.(\ref{11}):
\begin{eqnarray}
|\Phi_{out}^{(4)}\rangle &=&(1/\sqrt{2})\Big[(\hat b_1^{\dag}+\hat
b_2^{\dag})/\sqrt{2}\Big]^2(1/\sqrt{2})\Big[(\hat b_2^{\dag}-\hat
b_1^{\dag})/\sqrt{2}\Big]^2|0\rangle \cr &=&
\sqrt{3/8}\Big(|4_1,0_2\rangle+|0_1,4_2\rangle\Big) -
(1/2)|2_1,2_2\rangle.\label{15}
\end{eqnarray}

But in this section, we are interested in the
$|4_1,0_2\rangle$-term. From Eq.(\ref{15}), we find the
probability for this is $P_4(4,0)=3/8$. However, if the four
photons were classical particles, the classical probability is
simply $P_4^{cl}(4,0)=T^2R^2=1/16$. So the ratio between quantum
and classical prediction is $P_4^{qu}(4,0)/P_4^{cl}(4,0)=6$. The
six-fold ratio is a result of four-photon bunching effect.
Experimentally, it was first demonstrated by Ou {\it et
al.}\cite{ou99} The result is shown in Fig.7b.

The pair bunching effect is a result of constructive four-photon
interference. As shown in Fig.8, there are six possible ways to
arrange two pairs of photons (black and white circles) in four
detectors. If the four photons are classical particles, the black
and white pairs are distinguishable and we add the probabilities
of these six possibilities, resulting in $P_4^{cl}(4,0)=6|A|^2$
with $A$ as the probability amplitude for each possibility. On the
other hand, the four photons are quantum particles, there is no
way to distinguish between the black and white pairs. We then add
the amplitudes of each possibilities before obtaining the overall
probability $P_4^{cl}(4,0)=|6A|^2=6P_4^{cl}(4,0)$. The phases for
all six possibilities are same due to the same path for the four
photons. This leads to constructive four-photon interference. Note
that this is similar to the two-photon bunching effect in Sect.1.3
except that there are four detectors here so that there are more
possibilities.

\begin{figure}[bt]
\centerline{\psfig{file=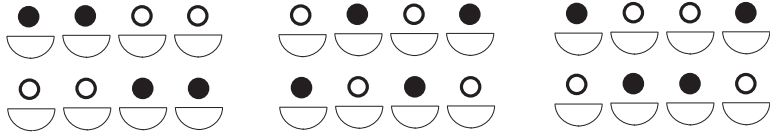,width=4in}} \vspace*{8pt}
\caption{Six possibilities for detecting two pairs of photons in
four detectors.}
\end{figure}

\subsection{Stimulated emission as a generalized photon bunching effect}

As is well-known, stimulated emission was first proposed by
Einstein\cite{ein} to explain the spectrum of blackbody radiation.
Phenomenologically, when a single photon interacts with an excited
atom, it can stimulate the atom to emit. The atom can of course
emit a photon spontaneously. From Einstein's $A$- and $B$-
coefficients, the rates of the stimulated emission and spontaneous
emission are same and are denoted as $R$. The overall rate is then
$2R$. When there are $N$ input photons, each photon may stimulate
the atom and the overall rate is then $(N+1)R$.

The enhancement effect due to stimulated emission can be used to
explain the photon bunching effect discussed in Sect.1.3. As shown
in Fig.9, the two photons detected in two-photon coincidence
measurement are from two excited atoms. In Fig.9a, the atoms in
excited state independently emit photons due to spontaneous
emission and two-photon detection probability in this case is
simply the product of individual emission probability:
$P_2^{sp}=P_1^2=R$. In Fig.9b, the detected two photons are from
stimulated emission, i.e., the photon spontaneously emitted from
one atom stimulates the emission of another atom. Since the rate
of stimulated emission is the same as that of spontaneous
emission, we have $P_2^{st}=R=P_2^{sp}=P_1^2$ so that the overall
probability is
\begin{eqnarray}
P_2= P_2^{st}+P_2^{sp}=2P_1^2,\label{16}
\end{eqnarray}
which is exactly the ratio of photon bunching effect.

\begin{figure}[bt]
\centerline{\psfig{file=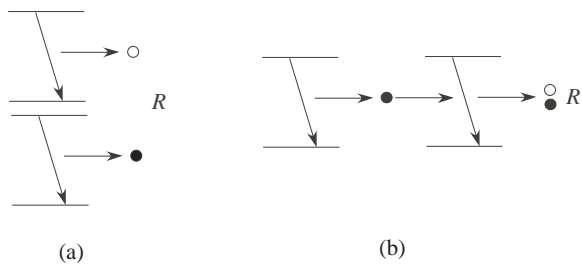,width=3.0in}} \vspace*{8pt}
\caption{Photon bunching effect as a result of stimulated
emission: two-photon coincidence from (a) spontaneous emission and
(b) stimulated emission.}
\end{figure}

Although Einstein did not give any argument for stimulated
emission, the quantum theory of light developed later fully
explained it. In essence, it is from the Bosonic relation of $\hat
a^{\dag}|N\rangle = \sqrt{N+1}|N+1\rangle$. But this explanation
relies on some complicated operator algebra. So is there a simpler
physical principle underlying the phenomenon of stimulated
emission?

As discussed in Sect.1.3, the photon bunching effect can be
understood as a constructive two-photon interference effect.
Combining this with the above explanation in terms of stimulated
emission, we may conclude that stimulated emission can also be
explained by multi-photon interference.

\begin{figure}[bt]
\centerline{\psfig{file=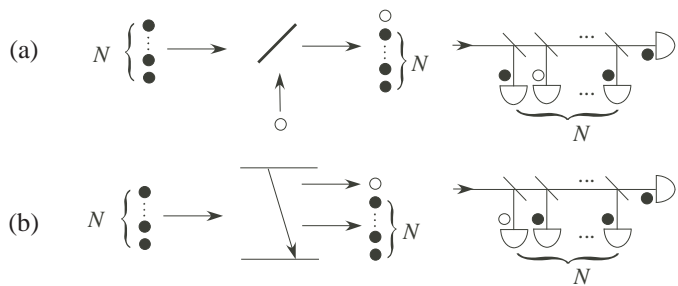,width=3.5in}} \vspace*{8pt}
\caption{Comparing (a) multi-photon interference to  (b)
stimulated emission by $N$ photons.}
\end{figure}

To see the connection, we look at the two schemes in Fig.10. For
the $(N+1)$-photon interference scheme in Fig.10a, it can be
shown, in a similar way to derive Eq.(\ref{15}), that the
$(N+1)$-photon detection probability is $P_{N+1}=(N+1)/2^{N+1}$,
which is $N+1$ times the probability $P_{N+1}^{cl} = 1/2^{N+1}$
when the $N+1$ photons were classical particles. The enhancement
factor is exactly the same as the stimulated emission scheme in
Fig.10b. This connection between stimulated emission and
multi-photon interference was recently demonstrated by Sun et
al.\cite{sun3}

\section{Generalized Hong-Ou-Mandel Effect and Destructive Multi-Photon Interference}

Although the two-photon Hong-Ou-Mandel effect cannot be
generalized to the case of more photons with a symmetric beam
splitter, as shown in Sect.2.1, we may consider its generalization
with an asymmetric beam splitter with $T\ne R$.

\subsection{Three-photon Wang-Kobayashi interferometer}

The three-photon generalization of Hong-Ou-Mandel interferometer
was proposed by Wang and Kobayashi\cite{wan}, who considered an
input state of $|2_1,1_2\rangle$ to a beam splitter with $T\ne R$.
With the method leading to Eqs.(\ref{11}, \ref{15}), we may easily
express the output state as
\begin{eqnarray}
&&|\Phi_{out}^{(3)}\rangle = \sqrt{3T^2R}~|3_1, 0_2\rangle +
\sqrt{3TR^2}~|0_1, 3_2\rangle \cr&&\hskip 1in+ \sqrt{T}(T-2R)|2_1,
1_2\rangle + \sqrt{R}(R-2T)|1_1, 2_2\rangle.\label{17}
\end{eqnarray}
Note that $P_3(2,1) = T(T-2R)^2$ and is equal to zero when
$T=2R=2/3$. Under this condition, Eq.(\ref{17}) becomes
\begin{eqnarray}
|\Phi_{out}^{(3)}\rangle = \Big(2~|3_1, 0_2\rangle +\sqrt{2}~
|0_1, 3_2\rangle  -\sqrt{3}~|1_1, 2_2\rangle\Big)\big/3.\label{18}
\end{eqnarray}

\begin{figure}[bt]
\centerline{\psfig{file=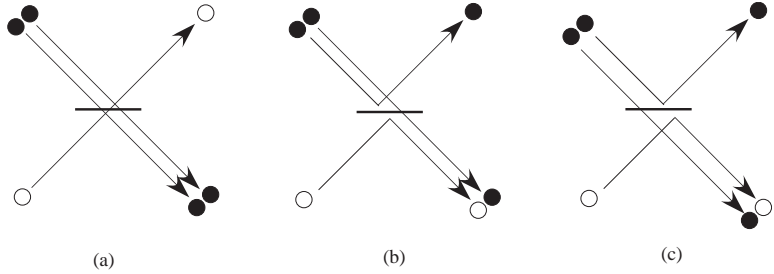,width=4in}} \vspace*{8pt}
\caption{Three possibilities for output of $|2_1,1_2\rangle$: (a)
all photons transmit; (b) and (c) one of the two photons from side
1 and the single photon from side 2 reflect.}
\end{figure}

Note that the disappearance of the $|2_1,1_2\rangle$ is due to
destructive three-photon interference. This complete
 cancellation of probability amplitude for the output of
 $|2_1,1_2\rangle$ is similar to the two-photon
Hong-Ou-Mandel effect and can be easily understood with the
picture in Fig.11, where there are three possible ways to obtain
an output of $|2_1,1_2\rangle$: (a) all three photons transmit
through the beam splitter with a probability amplitude of
$(\sqrt{2/3})^2$; (b) the single photon from side 2  and one of
the two photons from side 1 are reflected with a probability
amplitude of $-\sqrt{2/3}(\sqrt{1/3})^2$; and (c) the single
photon from side 2  and the other one of the two photons from side
1 are reflected with a probability amplitude of
$-\sqrt{2/3}(\sqrt{1/3})^2$. Thus the overall amplitude is
$(\sqrt{2/3})^2-2\sqrt{2/3}(\sqrt{1/3})^2 = 0$.

This generalized Hong-Ou-Mandel three-photon interference effect
was first demonstrated by Sanaka {\it et al.}\cite{san04} in
realizing a nonlinear phase gate.

\subsection{Fock state filtering effect}

The destructive three-photon interference effect in previous
section can be easily generalized to an arbitrary input state of
$|N_1,1_2\rangle$. With a beam splitter of $T, R$, we can easily
find the probability amplitude for an output of $|N_1,1_2\rangle$:
\begin{eqnarray}
A_{N+1}(N, 1)=\sqrt{T^{N-1}}(T-NR),\label{19}
\end{eqnarray}
which leads to the probability $P_{N+1}(N, 1)=T^{N-1}(T-NR)^2$,
which equals to zero when $T=NR=N/(N+1)$.

This effect can be used as a Fock state filter when conditioned on
the single-photon output at port 2. This idea was first proposed
and demonstrated by Sanaka {\it et al.}\cite{san06} and recently
put into application for generating an entangled photon
state\cite{res-2}. Consider an arbitrary state
$|\phi_{in}\rangle_1 = \sum_n c_n|n\rangle_1$ input at port one
and a single-photon state at port 2. The single photon at port 2
is often called the ancilla photon. We find from Eq.(\ref{19})
that when conditioned on the detection of a single photon at
output port 2, the output state at port 1 then becomes
\begin{eqnarray}
|\phi'_{out}\rangle_1 = {\cal N}\sum_n
c_n\sqrt{T^{n-1}}(T-nR)|n\rangle_1,\label{20}
\end{eqnarray}
where ${\cal N}$ is the normalization factor. Notice that when we
choose $T, R$ so that $R/T=n_0=$ integer, the state $|n_0\rangle$
disappears from the conditioned output state
$|\phi'_{out}\rangle_1$. Thus, the Fock state $|n_0\rangle$ is
filtered out.

\subsection{Two-Pair Wang-Kobayashi interferometer and its generalization}

The idea in previous sections can be applied to two pairs of
photons entering a beam splitter. For a beam splitter with
arbitrary $T, R$ and an input state of $|2_1,2_2\rangle$, we may
use the method that leads to Eqs.(\ref{11}) and (\ref{17}) to
obtain the output state as
\begin{eqnarray}
|\Phi_{out}^{(4)}\rangle &=& \sqrt {6 }TR \Big(|4_1, 0_2\rangle +
|0_1, 4_2\rangle\Big) + \Big[(T-R)^2-2TR\Big]|2_1, 2_2\rangle \cr
&&\hskip 1in + \sqrt{6TR}(T-R)\Big(|3_1, 1_2\rangle - |1_1,
3_2\rangle\Big).\label{21}
\end{eqnarray}
Obviously when $T=R=1/2$, Eq.(\ref{21}) becomes Eq.(\ref{15}). But
when $(T-R)^2-2TR=0$, or $T=(3\pm\sqrt{3})/6$ and
$R=(3\mp\sqrt{3})/6$, the $|2_1,2_2\rangle$ term disappears from
Eq.(\ref{21}), or $P_4(2,2) = 0$. Again, this disappearance of
$|2_1,2_2\rangle$ is a result of four-photon destructive
interference, similar to the Hong-Ou-Mandel effect.
Experimentally, this effect was first observed by Liu {\it et
al.}\cite{liu}

\section{Multi-Photon de Broglie Wavelength and Phase-dependent Multi-Photon Interference}

The interference effects discussed in Sects. 2 and 3 are phase
independent. They are either constructive with 0$^{\circ}$ phase
difference or destructive with 180$^{\circ}$ phase difference. In
this section, we will allow the phase difference to change and
reveal the more traditional interference pattern. In multi-photon
interference, phases of individual photons may be adjusted
independently and if the photons are spatially separated, we will
have nonlocal phase correlation that exhibits some dramatic
violation of local realism \cite{ghz}. On the other hand, when all
the photons are inseparable, they will sense the same phase shift
$\varphi$ resulting in an overall phase shift of $N\varphi$ with
$N$ as the total photon number. This corresponds to the case when
the $N$ photons form one entity and have an equivalent de Broglie
wavelength of $\lambda/N$ for the $N$-photon composite system,
with $\lambda$ as the single photon wavelength. Note that the key
in this system is that all the photons stick together and behave
like one single entity. For this type of states, if the photon
number is large, the states will become a Schr\"odinger cat state,
which is a superposition of macroscopic states.

\begin{figure}[bt]
\centerline{\psfig{file=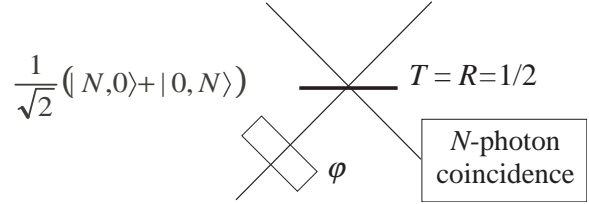,width=3.in}} \vspace*{8pt}
\caption{$N$-photon interferometry involving a NOON state.}
\end{figure}

\subsection{NOON state and Heisenberg limit}

One of these composite systems is the maximally entangled photon
number state of two modes\cite{ou97,bot,kok,lee,kok2}, which is
also called the NOON state and has the form of
\begin{eqnarray}
|NOON\rangle =\Big(|N_1, 0_2\rangle + |0_1,
N_2\rangle\Big)/\sqrt{2}.\label{22}
\end{eqnarray}
If we combine the two modes with a beam splitter and observe
$N$-photon coincidence measurement (Fig.12), it is straightforward
to show that the $N$-photon coincidence is proportional to
\begin{eqnarray}
P_N= (1/2)(1+\cos N\varphi).\label{23}
\end{eqnarray}
The phase dependence in Eq.(\ref{23}) may lead to a phase
measurement precision of $1/N$, or the so-called Heisenberg
limit\cite{hei}, which is the ultimate quantum limit of phase
measurement\cite{ou97}. In the following, we will see how to
produce a phase dependence in Eq.(\ref{23}) by quantum
interference.

The NOON state is a special kind of superposition states with a
total photon number as $N$. It lacks the middle terms such as
$|(N-1)_1, 1_2\rangle, |(N-2)_1, 2_2\rangle$, etc. Otherwise,
there would be terms of the form of $\cos m\varphi$ with $m < N$
in Eq.(\ref{23}). This would not lead to Heisenberg limit in phase
measurement.

\subsection{Generation of NOON state by quantum interference}

A NOON state cannot be generated by simply sending a Fock state of
photon number $N$ to a 50:50 beam splitter since the output state
contains all of the states mentioned above. On the other hand,
Hong-Ou-Mandel interferometer\cite{hom} provides method of how to
produce a NOON state. From Eq.(\ref{11}), we find that when
$T=R=1/2$, the output state from the beam splitter becomes
\begin{eqnarray}
|\Phi_2\rangle_{out} = \big(|2_1, 0_2\rangle-|0_1,
2_2\rangle\big)/\sqrt{2},\label{24}
\end{eqnarray}
which is a two-photon NOON state. The disappearance of the
$|1,1\rangle$ term is a result of two-photon interference. For the
three- and four-photon cases, however, we see from Eqs.(\ref{17})
and (\ref{21}) that only one middle term can be set to zero and
there still exist other unwanted middle terms.

\begin{figure}[bt]
\centerline{\psfig{file=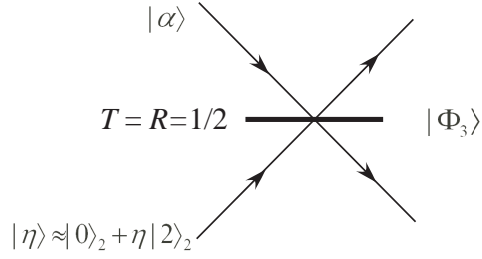,width=2.5in}} \vspace*{8pt}
\caption{Interference between a weak coherent state and a
two-photon state for the generation of a three-photon NOON state.}
\end{figure}

To introduce extra degrees of freedom for cancellation of all the
unwanted terms, we may use interference between a coherent state
and a two-photon state\cite{sso}. Consider the scheme in Fig.13
where a weak coherent state $|\alpha\rangle_1 \approx |0\rangle_1
+ \alpha |1\rangle_1 + (\alpha^2/\sqrt{2})|2\rangle_1 +
(\alpha^3/\sqrt{6})|3\rangle_1 +...$ and a two-photon state
$|\eta\rangle_2 \approx |0\rangle_2 +\eta|2\rangle_2 +...$ from
spontaneous parametric down-conversion enter a 50:50 beam splitter
from separate sides. Projecting to three-photon state, the input
state is $(\alpha^3/\sqrt{6})|3_1, 0_2\rangle  + \eta \alpha |1_1,
2_2\rangle$. With the method leading to Eqs.(\ref{11}) and
(\ref{17}), we find the output state as
\begin{eqnarray}
|\Phi_3\rangle_{out} = {(\alpha^2-\eta\sqrt{2})\alpha\over4}
\big(|2,1\rangle+|1,2\rangle \big)+
{\big(\alpha^2+3\eta\sqrt{2})\alpha\over4\sqrt{3}}
\big(|3,0\rangle+|0,3\rangle \big). \label{25}
\end{eqnarray}
When $\alpha^2=\eta\sqrt{2}$, the coefficient of $|2,1\rangle$ and
$|1,2\rangle$ are zero. The output state becomes a NOON state of
three photons.

This method has been generalized to four-, five-, six-, and
seven-photon cases\cite{liub} by introducing more degrees of
freedom. But the scheme becomes more and more complicated as
photon number gets large.  In 2004, Hofmann\cite{hof} proposed an
ingenious multi-photon interference method to cancel all the
unwanted middle terms mentioned above for the production of the
NOON state from $N$ independent single-photon states. The idea is
based on the algebraic identity:
\begin{eqnarray}
\prod_{n=1}^{N} \big(x-ye^{i\delta_n}\big)= x^N-y^N,\label{26}
\end{eqnarray}
where $\delta_n=2\pi( n-1)/N$.

\begin{figure}[bt]
\centerline{\psfig{file=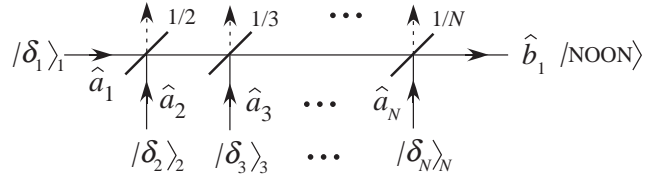,width=3.5in}} \vspace*{8pt}
\caption{Interference between a weak coherent state and a
two-photon state for the generation of a three-photon NOON state.}
\end{figure}

Consider the scheme in Fig.14 to merge $N$ modes into one with
$N-1$ beam splitters. Each mode is in a single-photon state with a
slight polarization twist:
\begin{eqnarray}
|\delta_n\rangle_n = {1\over \sqrt{2}} \big(|H\rangle_n -
e^{i\delta_n} |V\rangle_n\big)={1\over \sqrt{2}} \big(\hat
a_{Hn}^{\dag}- e^{i\delta_n} \hat a_{Vn}^{\dag}\big)|vac\rangle,
\label{27}
\end{eqnarray}
where $H,V$ denote the horizontal and vertical polarizations,
respectively. By generalizing the method that leads to
Eqs.(\ref{11}) and (\ref{17}) to $N-1$ beam splitter, we derive
the output state, which is quite complicated. However, if we are
only interested in the case when all $N$ photons are at the output
port 1, the projected state can be found as
\begin{eqnarray}
|\Phi_N(b_1)\rangle_{out} &=& {1\over (2N)^{N/2}}\prod_n(\hat
b_{H1}^{\dag} - e^{i\delta_n}\hat b_{V1}^{\dag})|vac\rangle \cr
&=& \sqrt{N!\over (2N)^{N}}\Big(|N_H, 0_V\rangle_1 - |0_H,
N_V\rangle_1\Big), \label{28}
\end{eqnarray}
where we used the identity in Eq.(\ref{26}). Note that the state
in Eq.(\ref{28}) is not normalized due to projection and it norm
$|||\Phi_N(b_1)\rangle_{out}||^2= 2 (N!)/ (2N)^{N}$ gives the
projection probability.

This scheme was implemented by Mitchell et al.\cite{mit} to
produce a three-photon NOON state.

\subsection{Demonstration of multi-photon de Broglie wavelength by projection measurement}

The starting point for Hofmann's scheme is single-photon state.
However, current technology is not mature enough to produce
single-photon states with consistent temporal profile\cite{yama}.
In this section, we will see how to demonstrate multi-photon de
Broglie wavelength without the need of NOON states. The idea is to
use some proper measurement schemes for projecting out the
unwanted terms in realization of NOON states.

\begin{figure}[bt]
\centerline{\psfig{file=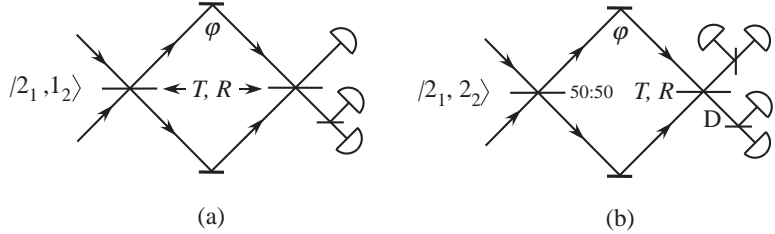,width=4.0in}} \vspace*{8pt}
\caption{Projection method for demonstrations of de Broglie
wavelength of (a) three photons and (b) four photons.}
\end{figure}

First consider the scheme in Fig.15a, which  was first proposed by
Wang and Kobayashi\cite{wan}. It uses twice the three-photon
Wang-Kobayashi interferometer, which was discussed in Sect.3.1.
The first beam splitter takes out the $|2_1,1_2\rangle$ term while
the second beam splitter together with three detectors removes the
$|1_1,2_2\rangle$ term, leaving only the contribution from the
NOON state part in Eq.(\ref{18}). Wang and Kobayashi\cite{wan}
proved that with $T=2/3$ for both beam splitters in Fig.15a, the
three-photon coincidence rate is indeed proportional to $1+\cos
3\varphi$, demonstrating three-photon de Broglie wavelength.
Experimentally, this scheme is implemented by Liu {\it et
al.}\cite{liu2}

Extension to four-photon case is straightforward: with an input
state of $|2_1,2_2\rangle$, we use choose $T=(3\pm\sqrt{3})/6$ for
the first beam splitter in Fig.15b to get rid of the $|2,2\rangle$
state and use a symmetric beam splitter as the second one. The
four-photon coincidence measurement of $P_4(2,2)$ at the output of
the interferometer projects out the $|3,1\rangle$ and
$|1,3\rangle$ terms as seen from Eq.(\ref{15}), leaving only the
NOON state contribution. Experimental implementation of the
four-photon case was performed by Liu {\it et al.}\cite{liu} with
a demonstration of a four-photon de Broglie wavelength.

An even simpler interferometer with four-photon de Broglie
wavelength can be formed with two symmetric beam splitters in the
scheme of Fig.15b. In order to eliminate the contribution of the
$|2_1,2_2\rangle$ term in Eq.(\ref{15}) with a second symmetric
beam splitter in Fig.15b, we make a detection of $P_4(3,1)$ in the
output of the interferometer. Because of the symmetric beam
splitter, $|2_1,2_2\rangle$ won't contribute to $P_4(3,1)$,
leaving only the contributions from the NOON state part in
Eq.(\ref{15}). This scheme was recently implemented by Nagata et
al.\cite{na}

Although extension of the above simple scheme to six-photon case
is possible, generalization to other photon numbers is difficult.
However, a scheme proposed by Sun {\it et al.}\cite{sun1} can be
easily generalized to arbitrary photon number. This scheme is a
reverse process of the scheme proposed by Hofmann\cite{hof}.

The scheme is sketched in Fig.16, where the input state is of an
arbitrary form:
\begin{eqnarray}
|\Psi_N\rangle_{in} =\sum_{n=0}^N c_n|N-n\rangle_H|n\rangle_V.
\label{29}
\end{eqnarray}
The fields at the detectors are related to the input fields as
\begin{eqnarray}
\hat b_n = (\hat a_H-\hat a_Ve^{i\delta_n})/N\sqrt{2} + ...,
\label{30}
\end{eqnarray}
where we omit the fields in vacuum. The $N$-photon coincidence
rate from all the detectors is proportional to
\begin{eqnarray}
P_N &\propto& \bigg\langle \prod_{n=1}^N\hat
b_n^{\dag}\prod_{m=1}^N\hat b_m \bigg\rangle=\bigg\langle (\hat
a_H^{\dag N}-\hat a_V^{\dag N})(\hat a_H^N-\hat a_V^N)\bigg\rangle
\Big/2^NN^{2N}\cr & =&P_{\Psi}(NOON)\Big/ 2^{N-1}N^{2N},
\label{31}
\end{eqnarray}
where
\begin{eqnarray}
P_{\Psi}(NOON)= |\langle NOON|\Psi_N\rangle|^2 =
|c_0-c_N|^2\label{32}
\end{eqnarray}
is the NOON state projection probability. If there is a phase
shift of $\varphi$ introduced between H and V fields and the two
fields have equal strength, we have $c_N/c_0=e^{iN\varphi}$. Then
Eq.(\ref{31}) becomes
\begin{eqnarray}
P_N \propto |c_0|^2(1-\cos N\varphi)/2^{N-2}N^{2N}, \label{33}
\end{eqnarray}
showing $N$-photon de Broglie wavelength.

\begin{figure}[bt]
\centerline{\psfig{file=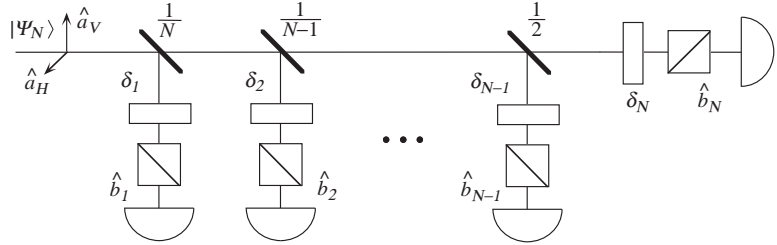,width=4in}} \caption{The scheme
of NOON state projection method for the demonstration of
$N$-photon de Broglie wavelength.}
\end{figure}

This NOON state projection scheme was applied to a four-photon
state from parametric down-conversion with four-photon de Broglie
wavelength\cite{sun2} and to  a weak coherent state with
six-photon de Broglie wavelength\cite{res}, respectively.

\section{Temporal Distinguishability of Photons}

Complementary principle of quantum mechanics states that
distinguishability inevitably degrades the interference effect.
For traditional single-photon interference, distinguishability may
only occur in paths. But for multi-photon interference,
distinguishability among different particles may also lead to
degradation of interference. Thus the multi-photon interference
effect discussed in the previous section can be used to
quantitatively characterize the degree of temporal
distinguishability of photons.

\subsection{Two-photon distinguishability and Hong-Ou-Mandel
interference}

As a matter of fact, the Hong-Ou-Mandel interference effect
discussed in Sect.1.2 depends on the overlap between the two
incoming photons. If the arrival times for the two photons at the
beam splitter is quite different, temporal distinguishability
between the two photons will diminish the interference effect.

Quantitatively, the visibility of the Hong-Ou-Mandel interference
effect is related to the two-photon spectral wave function
$\Phi_2(\omega_1,\omega_2)$ as\cite{wam}
\begin{eqnarray}
{\cal V}_2 = \int d\omega_1d\omega_2\Phi_2^*(\omega_1,\omega_2)
\Phi_2(\omega_2,\omega_1)\bigg/\int
d\omega_1d\omega_2|\Phi_2(\omega_1,\omega_2)|^2 , \label{34}
\end{eqnarray}
where the two-photon spectral wave function
$\Phi_2(\omega_1,\omega_2)$ is defined through an arbitrary
two-photon state of one dimension:
\begin{eqnarray}
|\Phi_2\rangle = \int d\omega_1d\omega_2\Phi_2(\omega_1,\omega_2)
\hat a^{\dag}_1(\omega_1)\hat a^{\dag}_2(\omega_2)|vac\rangle.
\label{35}
\end{eqnarray}
We omit the spatial degree of freedom for simplicity of
discussion. Its inclusion is straightforward.

From Eq.(\ref{34}), we find that ${\cal V}_2=1$, or the maximum
interference effect when
\begin{eqnarray}
\Phi_2(\omega_1,\omega_2) = \Phi_2(\omega_2,\omega_1), \label{36}
\end{eqnarray}
which is the condition for complete temporal indistinguishability
of the two photons. But ${\cal V}_2=0$, or no interference effect
when
\begin{eqnarray}
\int d\omega_1d\omega_2\Phi_2^*(\omega_1,\omega_2)
\Phi_2(\omega_2,\omega_1)= 0,\label{37}
\end{eqnarray}
which gives the criterion for complete temporal distinguishability
of the two photons. It turns out that Eqs.(\ref{36}, \ref{37}) can
be generalized to arbitrary number of photons

\subsection{Pair Distinguishability and Its Characterization}

The first generalization is to two pairs of photons by Ou {\it et
al.}\cite{ou99}, who considered a two-pair bunching effect
discussed in Sect.2.1. When the two pairs are well separated in
time, there is still some but less bunching effect. In fact, when
the two pairs are distinguishable, the input state becomes
\begin{eqnarray}
|\Phi_{in}^{(4)}\rangle ' =
|1_1,1_2\rangle\otimes|1'_1,1_2'\rangle,\label{38}
\end{eqnarray}
instead of $|\Phi_{in}^{(4)}\rangle  = |2_1,2_2\rangle$. The
output state is then
\begin{eqnarray}
|\Phi_{out}^{(4)}\rangle ' =(1/2)(|2_1,0_2\rangle-
|0_1,2_2\rangle)\otimes(|2'_1,0_2'\rangle-
|0'_1,2'_2\rangle),\label{39}
\end{eqnarray}
from which we find the probability $P_4(4,0) = 1/4$ and the ratio
to the classical probability is then
\begin{eqnarray}
P_4'(4,0)/P_4^{cl}(4,0) = 4.\label{40}
\end{eqnarray}
This value is reduced from the maximum value of 6 in Sect.2.1,
when the two pairs are indistinguishable from each other. Thus,
distinguishability results in degradation of the interference
effect.

For partial distinguishability of the pairs, Tsujino {\it et
al.}\cite{tsu} and de Riedmatten {\it et al.}\cite{de} attempted
to describe it as a mixed state between the two extreme cases
discussed above. However, this picture has some serious
problem\cite{ou05}. A better description is given in
Ref.\refcite{ou99-2} in terms of the quantity ${\cal E}/{\cal A}$
with
\begin{eqnarray}
{\cal A} \equiv \int
d\omega_1d\omega_1^{\prime}d\omega_2d\omega_2^{\prime}
|\Phi_2(\omega_1,\omega_2)\Phi_2(\omega_1^{\prime},\omega_2^{\prime})|^2,\label{41}
\end{eqnarray}
and
\begin{eqnarray}
{\cal E} \equiv \int
d\omega_1d\omega_1^{\prime}d\omega_2d\omega_2^{\prime}
\Phi_2(\omega_1,\omega_2)\Phi_2(\omega_1^{\prime},\omega_2^{\prime})
\Phi_2^*(\omega_1,\omega_2^{\prime})\Phi_2^*(\omega_1^{\prime},\omega_2),\label{42}
\end{eqnarray}
where $\Phi_2(\omega_1,\omega_2)$ is the two-photon wave function
defined in Eq.(\ref{35}).

For the two pairs of photons from parametric down-conversion, the
four-photon state is given by\cite{ou99-2}
\begin{eqnarray}
|\Phi_4\rangle = \int d\omega_1d\omega_2
d\omega_1'd\omega_2'\Phi_2(\omega_1,\omega_2)
\Phi_2(\omega_1^{\prime},\omega_2^{\prime})\hat
a^{\dag}_1(\omega_1)\hat
a^{\dag}_2(\omega_2)a^{\dag}_1(\omega_1')\hat
a^{\dag}_2(\omega_2')|vac\rangle. ~~~~\label{43}
\end{eqnarray}
So the four-photon wave function has the form of
\begin{eqnarray}
\Phi_4(\omega_1,\omega_2,
\omega_1',\omega_2')=\Phi_2(\omega_1,\omega_2)
\Phi_2(\omega_1^{\prime},\omega_2^{\prime}). \label{44}
\end{eqnarray}
Hence the condition for indistinguishable pairs, i.e., ${\cal
E}={\cal A}$, can be rewritten in terms of the four-photon wave
function as
\begin{eqnarray}
\Phi_4(\omega_1,\omega_2,
\omega_1',\omega_2')=\Phi_4(\omega_1',\omega_2,
\omega_1,\omega_2')= \Phi_4(\omega_1,\omega_2',
\omega_1',\omega_2). \label{45}
\end{eqnarray}
Note that the permutation is between primed and unprimed
variables, indicating permutation symmetry between two photons
with one from each pair. Thus we have pair exchange symmetry.

On the other hand, for the condition for complete distinguishable
pairs, i.e., ${\cal E}=0$, we have
\begin{eqnarray}
\int d\omega_1d\omega_1^{\prime}d\omega_2d\omega_2^{\prime}
\Phi_4(\omega_1,\omega_2,\omega_1^{\prime},\omega_2^{\prime})
\Phi_4^*(\omega_1,\omega_2^{\prime},\omega_1^{\prime},\omega_2)=0.\label{46}
\end{eqnarray}
Both Eqs.(\ref{45}) and (\ref{46}) are extension of Eqs.(\ref{36})
and (\ref{37}) to the four-photon case of two pairs. These can
further be generalized to arbitrary number of photons.

\subsection{Description of photon temporal distinguishability of an $N$-photon state}

With an $N$-photon state of arbitrary temporal profile in the form
of
\begin{eqnarray}
|\Phi_N\rangle = {\cal N}^{-1/2}\int
d\omega_1d\omega_2...d\omega_N \Phi(\omega_1, ..., \omega_N)\hat
a^{\dag}(\omega_1)\hat a^{\dag}(\omega_2)...\hat
a^{\dag}(\omega_N)|0\rangle,\label{47}
\end{eqnarray}
Eqs.(\ref{36}, \ref{37}, \ref{45}, \ref{46}) can be generalized to
\begin{eqnarray}
\Phi(\omega_1, ..., \omega_N) = \Phi(P_{ij}\{\omega_1, ...,
\omega_N\})\label{48}
\end{eqnarray}
for indistinguishability between two photons labelled as $i$ and
$j$ and
\begin{eqnarray}
\int d\omega_1d\omega_2...d\omega_N\Phi(\omega_1, ..., \omega_N) =
\Phi(P_{ij}\{\omega_1, ..., \omega_N\})=0\label{49}
\end{eqnarray}
for distinguishability between two photons labelled as $i$ and
$j$. Here $P_{ij}$ is the permutation operation between the
variables $\omega_i$ and $\omega_j$. ${\cal N}$ is the
normalization coefficient and takes the form of
\begin{eqnarray}
{\cal N} = \int d\omega_1d\omega_2...d\omega_N \Phi^*(\omega_1,
..., \omega_N)\sum_P \Phi(P\{\omega_1, ..., \omega_N\}).\label{50}
\end{eqnarray}
Obviously, ${\cal N}$ has a maximum value of $N!I$ with
\begin{eqnarray}
I = \int d\omega_1d\omega_2...d\omega_N |\Phi(\omega_1, ...,
\omega_N)|^2,\label{51}
\end{eqnarray}
when condition in Eq.(\ref{48}) is satisfied for all permutation.
It has the minimum value of $I$ when condition in Eq.(\ref{47})
applies to all permutation. The former corresponds to the case
when all the $N$ photons are indistinguishable of each other
(denoted as the $N\times 1$ case) whereas the latter to the case
when all the $N$ photons are well separated from each other and
become distinguishable (denoted as the $1\times N$ case).

Intermediate cases are described a combination of Eq.(\ref{50})
for some photons and Eq.(\ref{51}) for some other photons. For
example, the situation for two separate pairs (denoted as $2\times
2$ case) is described by Eq.(\ref{46}) for pair separation and by
Eq.(\ref{36}) for indistinguishability between photons within each
pair.

\subsection{Scheme of NOON state projection}

Since distinguishability in photons will influence the effect of
interference, we should be able to characterize the degree of
photon distinguishability through the measurement of the
visibility of interference, in a similar way as optical coherence
from the visibility of Young's double slit interference\cite{bw}.
But here we need $N$-photon interference because of the
involvement of $N$ photons. For example, the two-photon
Hong-Ou-Mandel effect provides a measure of temporal
distinguishability of two photons, as in Eq.(\ref{34}). The pair
bunching effect can be used to characterize the distinguishability
of two pairs of photons from parametric down-conversion, as in
Sect.5.2.

Among the schemes of multi-photon interference, the NOON state
projection measurement\cite{sun2}(Fig.16) easily applies to
arbitrary number of photons. Let us consider this scheme first.

From Eqs.(\ref{31}, \ref{32}), we find the $N$-photon coincidence
probability as
\begin{eqnarray}
P_N\propto |\langle NOON|\Psi_N\rangle|^2.\label{52}
\end{eqnarray}
For the input state of $|\Psi_N\rangle = |k, N-k\rangle ~(k\ne 0,
N)$, the coincidence probability is zero because of the
orthogonality of $\langle NOON|k, N-k\rangle =0 ~~(k\ne 0, N)$.

However, this is true only when all $N$ photons are
indistinguishable in time, due to complete destructive
interference. When photons are distinguishable, the interference
effect will deteriorate. The worst case is when the H-photons and
V-photons are completely separate and there is no interference at
all. The coincidence rate in this case sets a reference line to
compare with. As the H-photons and V-photons overlap in time, the
coincidence rate will drop below this reference, similar to the
Hong-Ou-Mandel effect. In this way,  we can define a visibility to
describe the interference effect. The size of the visibility will
depend on the degree of temporal distinguishability of the $N$
photons, thus providing a quantitative measure for the temporal
distinguishability of the $N$ photons.

The dependence of the visibility on the scenarios of temporal
distinguishability of the $N$ photons can be
calculated\cite{ou06}, based on the conditions in Eqs.(\ref{45},
\ref{46}). The form is very complicated in the NOON state
projection measurement. For the arbitrary case of $M$ photons in
H-polarization and $K$ photons in V-polarization, the visibility
was derived and given in Ref.\refcite{ou07}.

A simpler situation is for the input state of $|1_H, N_V\rangle$,
for which the visibility is given by
\begin{eqnarray}
{\cal V}_{N+1}^{NOON} = m/N, \label{53}
\end{eqnarray}
where $m$ is the number of V-photons that are indistinguishable
from the single H-photon. The other $N-m$ V-photons are completely
distinguishable from the $m+1$ photons. Thus by scanning the
relative delay between the single H-photon and the $N$ V-photons,
we may observe a number of dips, corresponding to the overlap
between the single H-photon and the groups of indistinguishable
V-photons. The visibility of the dips in Eq.(\ref{53}) gives the
number of V-photons in the corresponding group.

\begin{figure}[bt]
\centerline{\psfig{file=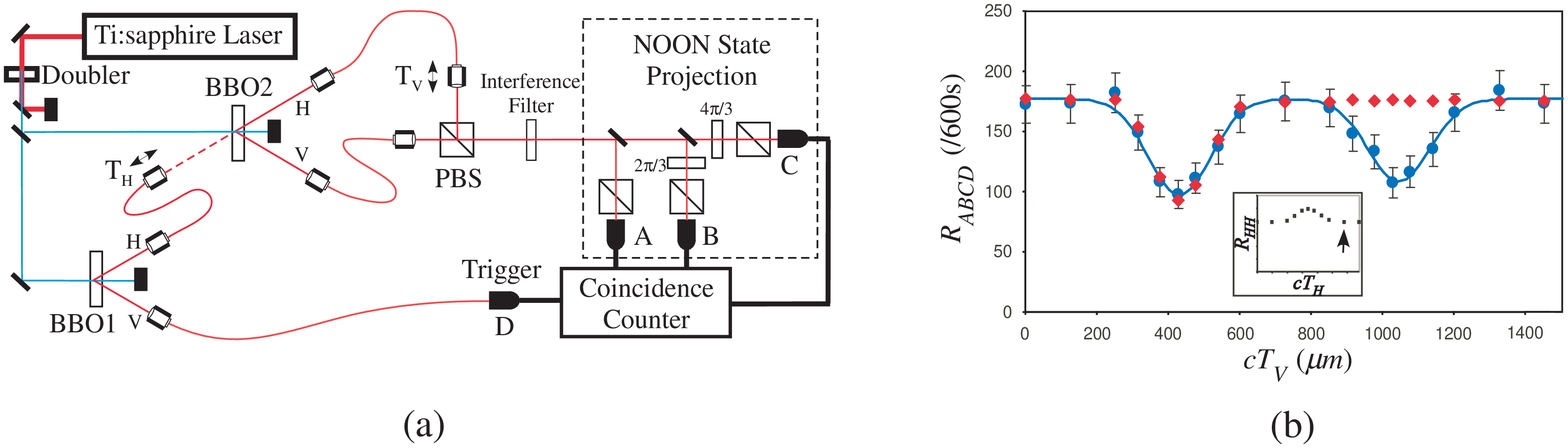,width=4.5in}} \vspace*{8pt}
\caption{(a) The schematics for demonstrating three-photon
temporal distinguishability; (b) the result of the experiment.
Reproduced with permission from Ref.[48].}
\end{figure}

The above situation was demonstrated by Liu {\it et
al.}\cite{liu06} for the case of $N=2$ with three photons from two
pairs of photons generated by two parametric down-conversion
processes. Fig.17 shows the setup and the results of the
experiment. Experimental demonstration of temporal
distinguishability for the input states of $|2_H, 2_V\rangle$ and
$|3_H, 3_V\rangle$ was performed by Xiang {\it et al.}\cite{xia06}

\subsection{Scheme of asymmetric beam splitter}

Besides the scheme of NOON state projection, we find from Sect.3
that another interference scheme with asymmetric beam splitter in
Sect.3.2 is a generalization of the Hong-Ou-Mandel interferometer.
This scheme can then be used to characterize the temporal
distinguishability of $N$ photons in a way similar to the NOON
state projection scheme.

For the simple situation of $|1_H, N_V\rangle$, we consider the
arrangement in Fig.18, which is an equivalent to the scheme with
an asymmetric beam splitter. The combination of a half wave plate
(HWP) and a polarization beam splitter (PBS) gives rise to a beam
splitter with variable transmissivity. For the scheme in Fig.18
and input state of $|1_H, N_V\rangle$, we find from Eq.(\ref{19})
that the coincidence probability of $(N+1)$ detectors is zero when
the rotation of the polarization due to the HWP is such that
$\sin^2 2\theta = 1/(N+1)$. $\theta$ is the angle of rotation of
the HWP. But this is true only when the $N+1$ photons are
indistinguishable. The coincidence probability is nonzero if part
of the $N+1$ photons become distinguishable. Coincidentally, it
can be shown\cite{ou07} that the visibility in the scheme of
Fig.18 is
\begin{eqnarray}
{\cal V}_{N+1}^{asy} = m/N, \label{54}
\end{eqnarray}
just like Eq.(\ref{53}) for the NOON state projection scheme.

\begin{figure}[bt]
\centerline{\psfig{file=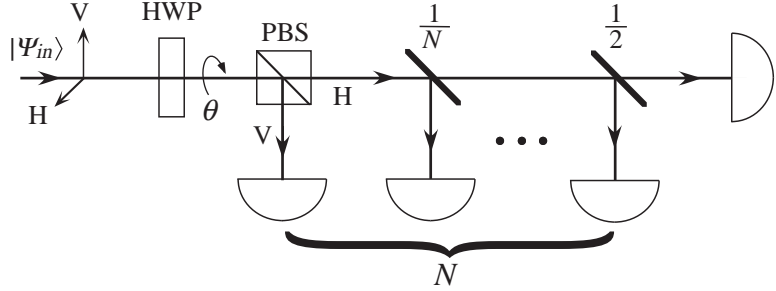,width=4.0in}} \vspace*{8pt}
\caption{The scheme with an asymmetric beam splitter for
characterizing the temporal distinguishability of $N$ photons.}
\end{figure}

However, the similarity ends right here. For other input state of
$|k_H, N_V\rangle$ with $k>1$, the visibility\cite{ou07} is even
more complicated than that of the NOON state projection scheme.
This scheme is implemented by Ou {\it et al.}\cite{liu07} for
$|1_H, 2_V\rangle$ state with a result similar to Fig.17b. The
experiment with an input state of $|2_H, 2_V\rangle$ in this
scheme was performed by Liu {\it et al.}\cite{liu06}

\section{Optical Coherence as Interpreted as Photon Indistinguishability}

Classical optical coherence theory, developed in the
1950s\cite{bw}, was based on second-order or single-photon
interference effect. In brief, the intensity distribution shows
the interference fringes pattern as
\begin{eqnarray}
I(x) \propto I_1+I_2+2\sqrt{I_1I_2}|\gamma| \cos 2\pi(x-x_0)/L,
\label{in}
\end{eqnarray}
where $I_1, I_2$ are the intensities of the two interfering
fields, $L$ is the fringe spacing along the $x$-direction, and
\begin{eqnarray}
\gamma \equiv \langle E_1^* E_2\rangle/\sqrt{I_1I_2} \label{ga}
\end{eqnarray}
is the degree of coherence between the two fields. Here $x_0$ in
Eq.(\ref{in}) is related to $Arg(\gamma)$.

Quantum coherence theory was later constructed by
Glauber\cite{gla} primarily along the same line as the classical
theory but with quantum formulism of operators and quantum states.
The physics was hidden beneath the complicated mathematical
formula.

As discussed in previous section, distinguishability of photons
lead to degradation of the visibility of interference. Thus the
two should be related somehow to each other. In the following, we
will make an initial attempt to reveal the connection.

In 1996, Javanaainen and Yoo\cite{ja} showed that in a single
realization, an interference fringe will form in the superposition
region of two groups of photons of the same number $N$,
respectively, i.e., with a state of $|N\rangle_1|N\rangle_2$.
Later, the study was extended by Ou and Su\cite{os} to the
superposition of two groups of photons with different photon
numbers $n$ and $m$, respectively, i.e., with a state of
$|n\rangle_1|m\rangle_2$. A quantum Monte Carlo
simulation\cite{os} shows that for the state of
$|n\rangle_1|m\rangle_2$, there is an interference fringe forming
with a probability distribution of
\begin{eqnarray}
P(x) \propto n+m+2\sqrt{nm} \cos 2\pi(x-x_0)/L, \label{55}
\end{eqnarray}
where $x_0$ is arbitrary and $L$ is the fringe spacing. If we
compare the above with Eq.(\ref{ga}), we find the normalized
degree of coherence is simply $\gamma = 1$. This is not surprising
in the sense that the photons in the quantum state $|n,m\rangle$
belong to one wave function and are all indistinguishable in the
superposition region.

On the hand, if there is partial indistinguishability among the
photons, from the discussion in previous section we find that the
visibility will drop. Assume that the input state is
$|N\rangle_1|M\rangle_2$ but only $n$ photons among the $N$
photons in mode 1 are indistinguishable from $m$ photons among the
$M$ photons in mode 2. Therefore, only the $n+m$ photons will give
rise to an interference pattern, described by Eq.(\ref{55}). The
rest photons, i.e., $N-n$ photons from mode 1 and $M-m$ photons
from mode 2, are distinguishable and produce no interference
fringe. Thus the probability distribution in this case is given by
\begin{eqnarray}
P'(x) &\propto &N-n +M-m + n+m+2\sqrt{nm} \cos 2\pi(x-x_0)/L \cr
&=& N +M+2\sqrt{nm} \cos 2\pi(x-x_0)/L. \label{56}
\end{eqnarray}
Comparing to Eq.(\ref{ga}), we have
\begin{eqnarray}
\gamma' = \sqrt{nm/NM} . \label{57}
\end{eqnarray}
Note that $n/N$ and $m/M$ are the percentages of indistinguishable
photons in the two groups, respectively. Thus the degree of
coherence is related to the percentage of indistinguishable
photons among the photons involving in interference.

\section{Summary and Discussion}

This paper has discussed various interference effects involving
multiple photons. Some are constructive interference effects
(photon bunching); some are destructive interference effects
(Hong-Ou-Mandel effect); and some are phase dependent interference
effects (multi-photon de Broglie wavelength). We find that we need
modify Dirac's statement on photon interference to understand
multi-photon interference effects. We also find that photon
distinguishability leads to degradation in interference effect,
confirming complementary principle of quantum mechanics in
quantitative manor.

Although we only discussed the problem of temporal
distinguishability between two pairs in one interference scheme in
Sect.5.2, it has been shown\cite{ou07} that all the interference
experiments involving two pairs of photons in Sects.2-4 can be
explained in terms of the quantity ${\cal E}/{\cal A}$, which
describes the temporal distinguishability between the two pairs.
This point can be generalized to all of the multi-photon
interference schemes. Further study\cite{ou07-2} shows that the
scheme with stimulated emission by $N$ photons can be used to
characterize quantitatively the temporal distinguishability of the
incoming $N$ photons, since it is a constructive multi-photon
interference effect.

\section*{Acknowledgements}

This paper is based on a lecture series in the summer school on
cold atoms and molecules and quantum optics, held in Eastern China
Normal University in August, 2007. The author is grateful to Prof.
Weiping Zhang for his invitation and support. The work described
in this paper was supported by the US National Science Foundation
under Grant No. 0245421 and No. 0427647. The author acknowledges
Prof. Guangcan Guo for his support via National Fundamental
Research Program of China and the Innovation Funds of Chinese
Academy of Sciences.

\end{document}